\documentclass[letterpaper, 10 pt, conference]{ieeeconf} 
\pdfoutput=1
\IEEEoverridecommandlockouts                            
\usepackage{graphicx}
\usepackage[implicit]{hyperref}

\title{\LARGE \bf
Resistive communications based on neuristors
}

\author{David Alejandro Trejo Pizzo$^{1}$
\thanks{*This work was not supported by any organization}
\thanks{$^{1}$David Alejandro Trejo Pizzo is with NyVind Research Labs and School of Engineering, Universidad de Palermo, C1175ABT Buenos Aires, Argentina
        {\tt\small dtrejopizzo@gmail.com}}%
}

\begin{document}

\maketitle
\thispagestyle{empty}
\pagestyle{empty}

\begin{abstract}

Memristors are passive elements that allow us to store information using a single element per bit. However, this is not the only utility of the memristor. Considering the physical chemical structure of the element used, the memristor can function at the same time as memory and as a communication unit. This paper presents a new approach to the use of the memristor and develops the concept of resistive communication.

\end{abstract}

\section{INTRODUCTION}

The memristor is the fundamental non-linear circuit element, with uses in computing and computer memory. Memristors are basically a fourth class of electrical circuit, joining the resistor, the capacitor, and the inductor, that exhibit their unique properties primarily at the nanoscale.

Theoretically, a memristor is a concatenation of a memory and a resistor that maintains a relationship between the time integrals of current and voltage across a two terminal element. Thus, its resistance varies according to a memristance function.

\begin{figure}[h]
\begin{center}
\includegraphics[scale=0.4]{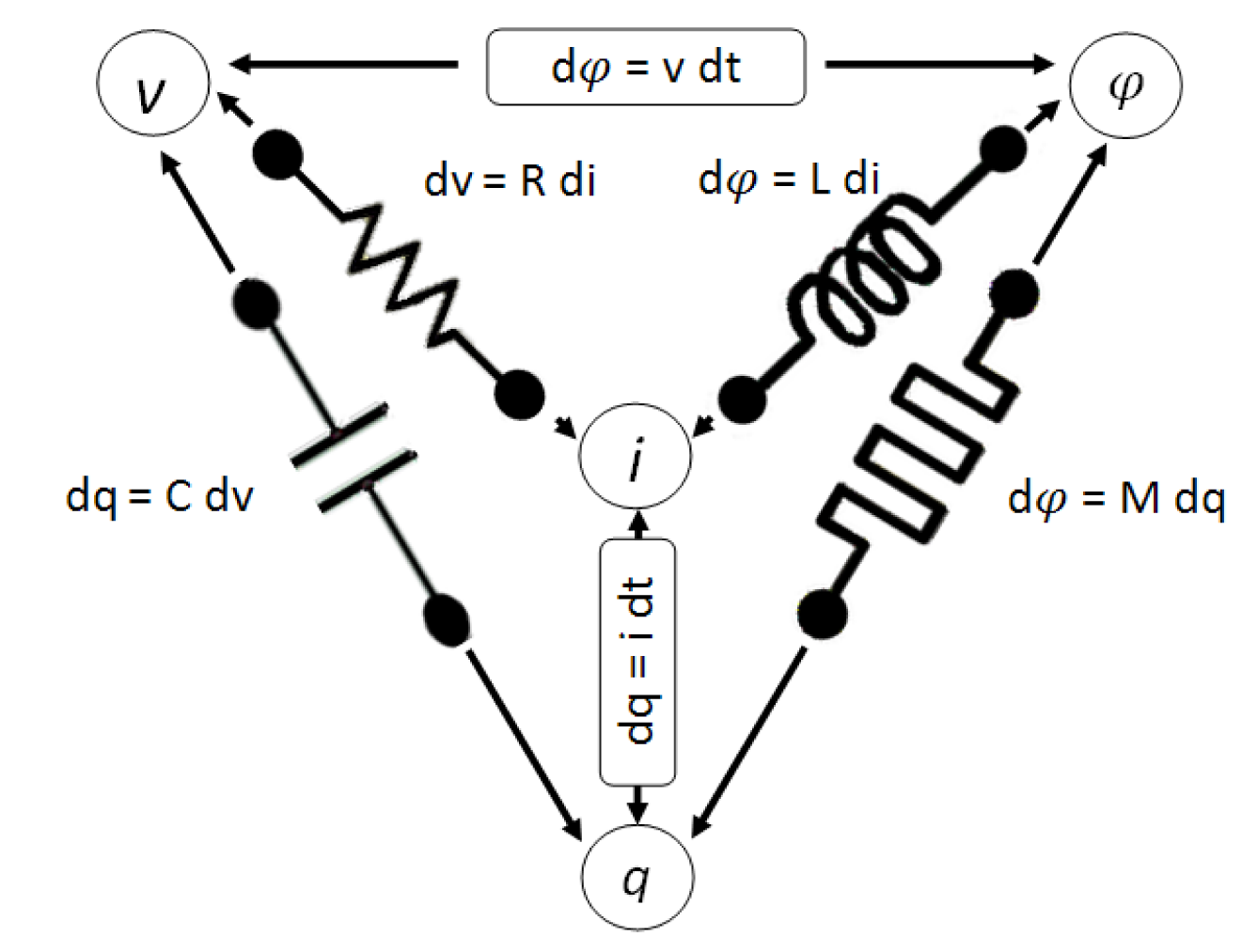}
\end{center}
\caption{Fundamental circuit elements}
\label{fig:elements}
\end{figure}

The material implementation of memristive effects can be determined in part by the presence of hysteresis (an accelerating rate of change as an object moves from one state to another) which, like many other non-linear anomalies turns out to be less an anomaly than a fundamental property of passive circuitry.

The original definition of the memristor is derived from circuit theory: besides resistor, capacitor and inductor, there must exist a fourth basic two-terminal element that uniquely defines the relationship between the magnetic flux $\phi$ and the electric charge $q$ passing through the device, or

$$d\phi = M * dq$$

\subsection{ReRAM}

ReRAM (Resistive Random Access Memory) is a resistive switching memory proposed as a nonvolatile memory. The phenomenon of resistive switching has been observed in a wide variety of materials, however, the mechanism responsible for the switching behavior seems to differ between them. On the basis of I-V characteristic curves, switching behaviors can be classified into two types, unipolar (non-polar) and bipolar.

In unipolar resistive switching, the switching process depends on the amplitude of the applied voltage, but not on the polarity. This type of switching behavior has been observed in many binary metal oxides such as SiO, NiO and CuO.

\begin{figure}[h]
\begin{center}
\includegraphics[scale=0.4]{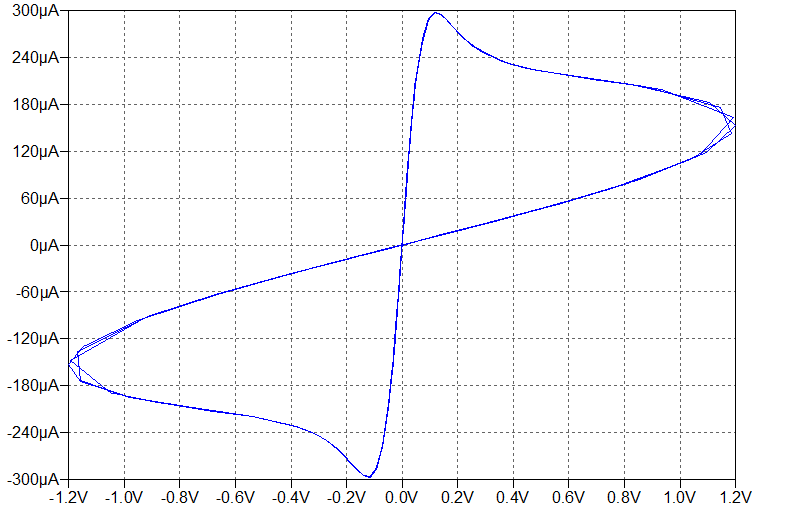}
\end{center}
\caption{I-V curve of a CuO memristor}
\label{fig:elements2}
\end{figure}

A memory cell initially in a virgin state, ie prior to the application of any voltage pulse, is in a highly resistive state. When the application of a voltage pulse is high enough, it is put into the device in a low resistance state (LRS). This process is called forming. The rupture value of the dielectric of the material tells us approximately at what range of voltages we should find this threshold voltage for the forming process. After the forming process, the cell from an LRS state is switched to a high resistance state (HRS) by applying a threshold voltage (reset process). Changing from an HRS state to an LRS state is achieved by applying a threshold voltage that is usually greater than the reset voltage, but less than the forming voltage. In the process of set and forming, the current is limited by the compliant current (Icc) of the control system or, more practically, by adding a series resistor acting as a voltage divider.

\section{MEMRISTOR CIRCUITS}

For almost fifty years, integrated electronic circuits built with semiconductor devices have provided significant growth in the number of processing elements and memory bits available to system developers.

\begin{figure}
\begin{center}
\includegraphics[scale=12]{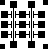}
\end{center}
\caption{Designend array of neuristors contacts}
\label{fig:contact}
\end{figure}

This growth has provided orders of magnitude improvements in speed, power consumption, and reliability, together with significant reductions in the cost per device but trends like this are direct consequences of frequent miniaturization in the semiconductor fabrication process. According to "Moore Law" this will reach an end soon. When device sizes are no longer scalable, microelectronic technology needs innovations to support novel applications and here comes the memristor.

\begin{figure}
\begin{center}
\includegraphics[scale=0.8]{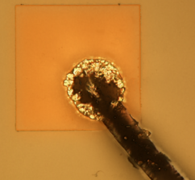}
\end{center}
\caption{Neuristor contact}
\label{fig:contact2}
\end{figure}

Memristors are used in hardware neural networks, both to implement different learning algorithms and back propagation. It is also used to build neuromorphic systems (hardware that mimics the brain). Memristors can also be used in analog circuits, for example as reconfigurable resistors to change the properties of the circuit. Another interesting application is the use of memristors as part of logic circuits. 

\section{RESISTIVE COMMUNICATIONS}

We know that neurons communicate with each other through the small spaces between them, in a process known as synaptic transmission (where synapses are the connections between neurons). Information goes from one cell to another by neurotransmitters such as glutamate, dopamine or serotonin, which activate the receptors in the receiving neuron to transmit excitatory or inhibitory messages.

Three main types of synaptic transmission are distinguished. The first two mechanisms constitute the main forces governing the neural circuits and can be replicated using memristor circuits, from now on called "neuristors":

\begin{itemize}
\item Exciting transmission: one that increases the possibility of producing an action potential;
\item Inhibitory transmission: that which reduces the possibility of producing an action potential;
\item Modulating transmission: that which changes the pattern and / or frequency of the activity produced by the cells involved.
\end{itemize}

\begin{figure}
\begin{center}
\includegraphics[scale=1]{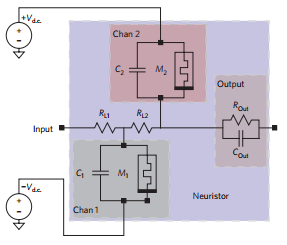}
\end{center}
\caption{Neuristor made with Mott memristors}
\label{fig:neuristor}
\end{figure}

As it is shown in "A scalable neuristor built with Mott memristors" and in figure \ref{fig:neuristor} a neuristor can be made with Mott memristors. The channels consist of Motts M1 and M2, each with a characteristic parallel capacitance
(C1 and C2, respectively) and are biased with opposite polarity d.c. voltage sources.

\subsection{The conduction mechanism}

To support the general function of the nervous system, neurons have evolved unique capabilities for intracellular signaling (communication within the cell) and intercellular signaling (communication between cells). To achieve long distance, rapid communication, neurons have evolved special abilities for sending electrical signals (action potentials) along axons. This mechanism, called conduction, is how the cell body of a neuron communicates with its own terminals via the axon. Communication between neurons is achieved at synapses by the process of neurotransmission.

\begin{figure}
\begin{center}
\includegraphics[scale=1]{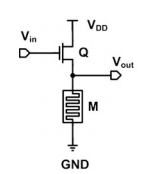}
\end{center}
\caption{Neuristor proposed circuit for synapses}
\label{fig:synapse}
\end{figure}

We can use the neuristor M to achieve this communication as it is shown in figure \ref{fig:synapse}.

To begin conduction, a potential is generated near the cell body portion of the axon, here the Vin of the Q transistor. But whereas an electrical signal in an electronic device occurs because electrons move along a wire, an electrical signal in a neuron occurs because ions move across the neuronal membrane. Ions are electrically charged particles. The protein membrane of a neuron acts as a barrier to ions. Ions move across the membrane through ion channels, here nanowires in a dielectric thin layer, that open and close due to the presence of neurotransmitters.

\begin{figure}
\begin{center}
\includegraphics[scale=1]{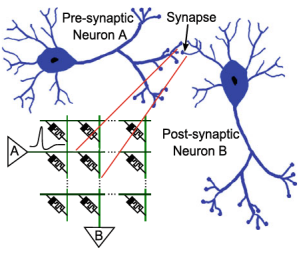}
\end{center}
\caption{Neuristor simplified grid}
\label{fig:synapsepro}
\end{figure}

When the concentration of ions on the inside of the neuron changes, the electrical property of the membrane itself changes. Normally, the membrane potential of a neuron rests as -70 millivolts (and the membrane is said to be polarized). The influx and outflux of ions (through ion channels during neurotransmission) will make the inside of the target neuron more positive (hence, de-polarized). When this depolarization reaches a point of no return called a Vthreshold, a large electrical signal is generated. This is the action potential.

This signal is then propagated along the axon (and not, say, back to its dendrites) until it reaches its axon terminals. An action potential travels along the axon quickly, moving at rates up to 150 meters (or roughly 500 feet) per second. Conduction ends at the axon terminals.

Axon terminals are where neurotransmission begins. Hence, it is at axon terminals where the neuron sends its Vout to other neurons. At electrical synapses, the Vout will be the electrical signal itself. Neuristors communicate with other neuristors with every set and reset action, changing the resistance level and forming nano wires that can save the memory of past memristance levels. This conducction mechanism is what I would like to call resistive communication.

\subsection{A nanoscale communication protocol}

Resistive communications is an approach to communicate between physical layers of neuristor with focus in creating neuromorphic circuits. To make this communication mechanism compatible with other nano devices it is essential to use a standard. A conceptual framework provides the organization and structure required to develop conceptual models of nanoscale communication. The IEEE Std 1906.1 Recommended Practice for Nanoscale and Molecular Communication Framework provides this precise, common definition of nanoscale communication and a general framework that balances concepts with broad applicability. This includes metrics, use-cases, and a reference model witch it is followed here.

\subsection{Reference model for molecular communications}

The core of the simulator has been also extended to model  molecular communications based on the pure diffusion process. the UML diagram of classes modeling the Molecular example. Also in this case, the diagram only reports the most important data members and functions, whereas some details about relationships among objects have been omitted.

It is assumed that molecules diffuse into the medium according to Brownian motion. In that hypothesis propagation of this pulse can be analytically modeled by Ficks laws of diffusion, which expresses the concentration of molecules as a function of distance and time. In particular, the molecular concentration at any point in space is expressed in Equation

$$c(r,t) = \frac{Q}{(4*\pi*D*t)^{3/2}}*e^{\frac{-r^{2}}{4*D*t}}$$

where
\begin{itemize}
\item c is the molecular concentration
\item r is the distance between sender and receiver
\item Q is the number of molecules released by the sender
\item D is the diffusion coefficient
\item t is the time variable
\end{itemize}

\begin{figure}
\begin{center}
\includegraphics[scale=0.5]{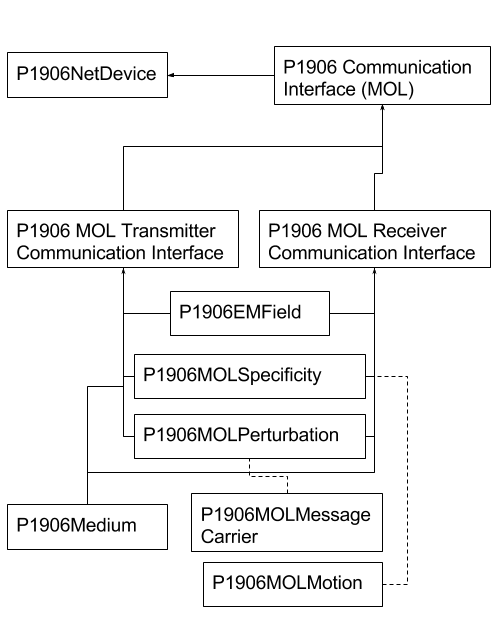}
\end{center}
\caption{IEEE P1906 reference model}
\label{fig:synapsepro2}
\end{figure}

\section{CONCLUSIONS}

The resistive communication presents a model on which to build neuromorphic networks that emulate the functioning of the electrical synapse and the communication of data between neurons. Much work remains to be done to build a neuromorphic circuit that emulates brain functions, but we are closer to achieving small neural networks to physically build pattern recognition algorithms. The memristor and the neuristor construction are the ideal elements for its physical characteristics to build neural networks and emulate the electrical synapse using the resistive switching phenomenon.

\addtolength{\textheight}{-12cm}


\end{document}